\documentclass[aps,twocolumn,superscriptaddress]{revtex4-1}
\usepackage{graphicx}
\begin{document}
\title{Emergence of communities on a coevolutive model of wealth interchange}
\author{A. Agreda}
\affiliation{Centro de F\'isica Fundamental. Facultad de Ciencias.  Universidad de Los Andes. M\'erida 05101, Venezuela.}
\author{K. Tucci}
\affiliation{Centro de F\'isica Fundamental. Facultad de Ciencias.  Universidad de Los Andes. M\'erida 05101, Venezuela.}
\affiliation{SUMA-CESIMO Universidad de Los Andes. M\'erida 05101, Venezuela.}



\begin{abstract}
We present a model in which we investigate the structure and evolution of a random network that connects agents capable of exchanging wealth. Economic interactions between neighbors can occur only if the difference between their wealth is less than a threshold value that defines the width of the economic classes. If the interchange of wealth cannot be done, agents are reconnected with another randomly selected agent, allowing the network to evolve in time. On each interaction there is a probability of favoring the poorer agent, simulating the action of the government. We measure the Gini index, having real world values attached to reality. Besides the network structure showed a very close connection with the economic dynamic of the system.
\end{abstract}
%
\maketitle
\section{Introduction}
\label{intro}

Within the modeling of collective behavior, 
a strong interest in the problem of
structure formation in networks of interacting agents has been developed,
such as in models of market, scattering of rumors, opinions formation, etc.
Networks consist of a number of nodes or agents (individuals, countries, firms of investors, banks, etc.) connected and related by links. The particular pattern of connections specifies the topology of the network, such connections can be established, removed or change its strength as the system evolves over time. When the structure of the network changes because of the dynamic of the nodes states and therefore there is a coupling between topology and states then it is a coevolutionary system or adaptive network.

Coevolutionary systems emerge in many different applications and have been studied in epidemic propagation \cite{epidemic1,epidemic2}, technical distribution networks \cite{technicalNet1,technicalNet2}, neural networks \cite{neural1,neural2}, models of social dynamics \cite{social1,social2,social3}, game theory \cite{game}, ecological research models \cite{eco1,eco2}, chemical networks \cite{chemical1,chemical2}.

This paper proposes a microscopic model of wealth exchange between agents
located on a dynamical network where the structure formation and wealth distribution are characterized and studied.
As coevolutionary system, the model can be classified using the general framework for systems with coevolution between topology and dynamics \cite{HCTG2011} as a system with rewiring process type DR, i.e. a model where the actions of disconnection and reconnection are respectively based on dissimilarity (D) and randomness (R) mechanisms.

Many works have been directed to formulating similar models to the proposed here: Pianegonda {\em et al.}~\cite{PI03} presented a model of patterns of redistribution of wealth on a one-dimensional network. 
Iglesias \emph{et al.} \cite{Iglesias} studied the distribution of wealth in an agent-based model with an element of risk aversion on each agent. 
Laguna \emph{et al.} \cite{Laguna} looked at the effect of social stratification in the distribution of wealth in a system of economic agents with interactions that are limited to only interact within the same economic class. 
Following this direction, Herrera \emph{et al.} \cite{JL,HCT2011} added to the concept of stratification of Laguna \emph{et al.} the concepts of neighborhood and spatial location. 
Also, there are some models \cite{IglesiasGoncalvesPianegonda,Garlaschelli} that have studied the influence of network topology on economic dynamics.

The proposed coevolutive model of wealth exchange is presented in Sec.~\ref{sec:TheModel}.
Results are shown in Sec.~\ref{sec:Results} where  the Gini index is used as an order parameter to characterize the wealth distribution and network structure is characterize through two order parameters: the fraction of agents on the largest component of the network and the network modularity.
The emergence of networks with communities are shown, and their relation with the Gini index is explained through a phase diagram.
The conclusions are presented in Sec.~\ref{sec:Discussion}.

\section{The model}
\label{sec:TheModel}

The model proposed consists of $N$ agents that can exchange wealth with their neighbors. The exchange of wealth is based on the model of interaction rules proposed by Herrera \emph{et al.}\cite{JL}.
Nevertheless, unlike this one, here the agents form a dynamic network, whose undirected links can be rewired over the time. 
Each agent $i = 1,2, \dots, N;$ is characterized at time $t$ by its wealth $w_i(t)$ and the set of its $k_i(t)$ neighbors  $\eta_i(t)$.
The initial value of wealth is $w_i(0) = 1, \ \forall \ i $ and its set of neighbors, $\eta_i(0)$, is obtained from the network, which at $t = 0$, is a random network type Erd\"{o}s-Renyi~\cite{ER59} with a degree $\bar{k}=N^{-1}\sum_i k_i(0)$. 
In addition, each agent has a risk aversion $ \beta_i$ that characterizes how much the fact that the agent $i$ is not willing to risk in an economic transaction, thus the fraction of wealth that the agent $i$ is willing to risk on each transaction is $(1 - \beta_i)$. For each simulation the values $ \beta_i \in [0,1] $ are distributed randomly and remain fixed during all the time. 

At each instant $t$, an agent $i$ is chosen randomly from the $N$ agents on the network. Then a second agent $j$ is chosen  from the neighborhood of $i$, i.e., 
$j \in \eta_i $, randomly as well. If the normalized wealth difference between them does not exceed a threshold $u$, that is
\begin{equation}
  \frac{|w_i(t)-w_j(t)|}{\max(w_i(t),w_j(t))}<u \;,
  \label{dif_riquezas}
\end{equation}
the wealth exchange is performed. Note that in this model the parameter $u$ measures the width of economic classes and the exchange of wealth can only occur between neighboring agents that belong to the same economic stratum.

But if the chosen agents $i$ and $j$ are not of the same economic class, i.e., the inequality of eq.(\ref{dif_riquezas}) is false, then a rewiring process is fired, disconnecting the agent $i$ from $j$ and connecting it with another randomly chosen agent, $j^*$, that was not in the neighborhood of $i$.
As the links of the network are undirected, when $i$ and $j$  are disconnected, the agent $j$ is taken out from the set $\eta_i$, as well as $i$ is also extracted from the set $\eta_j$. And by connecting $i$ to $j^*$ each agent is added to the neighbors set of the other.

For the wealth exchange it is established that no agent can gain more than the invested quantity, so the amount to be exchanged is given by
\begin{equation}
dw=\min\left[(1-\beta_i)w_i,(1-\beta_j)w_j\right] \;.
\label{dw}
\end{equation}
To emulate the public policies which aim is contribute with wealth redistribution, in the model there is a probability $p \geq 1/2$ to favour the poorer of the two interacting agents, defined as
\begin{equation}
p=\frac{1}{2}+f\times\frac{|w_i(t)-w_j(t)|}{w_i(t)+w_j(t)} \;,
\label{prob_maspobre}
\end{equation}
where $f$ is a parameter that ranges from $f = 0$, for an equal probability of favoring each agent, to $f = 1/2$, where the probability of favoring the poorer is maximum.
Thus, in each interaction the poorer agent has a probability $p$ to be favored and  increase $dw$ its wealth and the richer agent to lose this amount of its wealth, while $ (1-p) $ is the probability that otherwise happens.
In this way, the total wealth of the system is conserved, that is
$ W=\sum_{i=1}^{N}w_i(t)=\sum_{i=1}^{N}w_i(0)\; \forall i\;. $

\section{Results}
\label{sec:Results}

The results shown in this section were done with undirected networks of $N=10^4$ agents initially connected randomly. The degree of the network, i.e. the average number of neighbors per agent, is $\bar{k}=16$. The simulation time was $T=10^9$ iterations. Each point corresponds to the average value of 5 realizations.

Gini index is used as order parameter as a way to characterize the
statistical properties of the wealth distribution in the system.
This quantity measures the degree of inequality in an economic system and is given by
\begin{equation}
G=\frac{\sum_{i=1}^N \sum_{j=1}^N |w_i - w_j|}{2 W N} \;.
\label{eq:G}
\end{equation}
In a large population the Gini index can take values between 0 and 1. A fully equitable wealth distribution, where $w_i=w_j \; \forall \, i,j$; corresponds to $G=0$, while a totally unequal distribution, where one agent has all the richness of the system and the others have no wealth at all, gives a Gini index $G=1$.

\begin{figure}[htb]
\centerline{
\includegraphics[width=0.40\linewidth,angle=90]{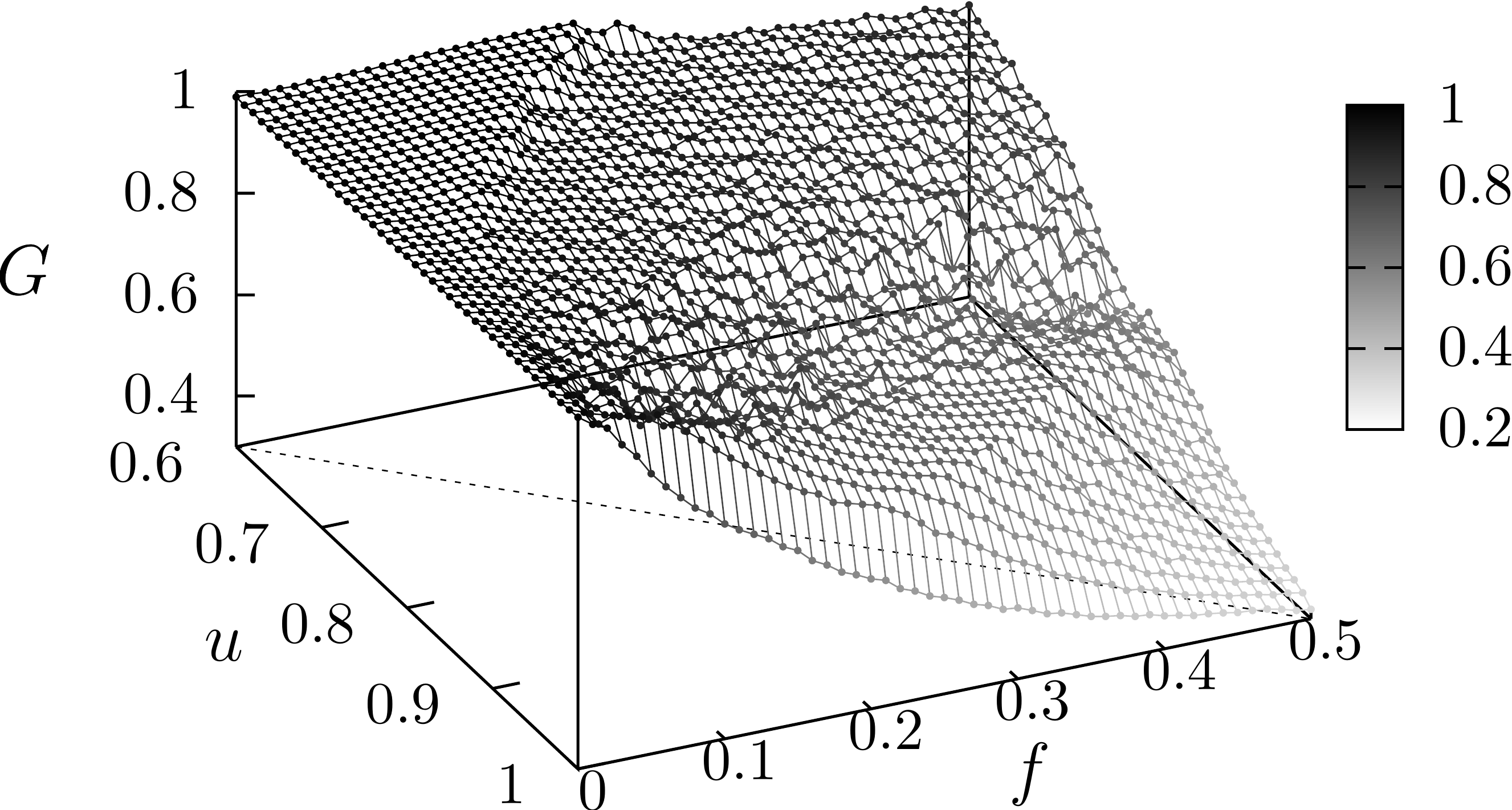}
}
\caption{Gini index $G$ as a function of the parameters $u$ and $f$. The parameter $u$ represents the width of the economic classes while the public politics of wealth redistribution are represented by $f$.}
\label{fig:G}
\end{figure}

Figure \ref{fig:G} shows the value of the Gini index in the parameters space ($u$;$f$). 
As might be expected, the Gini index reaches its maximum values when the parameter $f~=~0$, i.e. when there are no public politics of wealth redistribution. 
In this figure it is also appreciated that for each value of the parameter $u$ there are some values of the public politics of wealth redistribution for which the Gini index begins to decay, $f_G$. 
Note that the values of $f_G$ decrease while $u$ increases until reach $f_G=0$ when $u=1$.

\begin{figure}[h]
\centerline{
  \includegraphics[width=0.40\linewidth,angle=90]{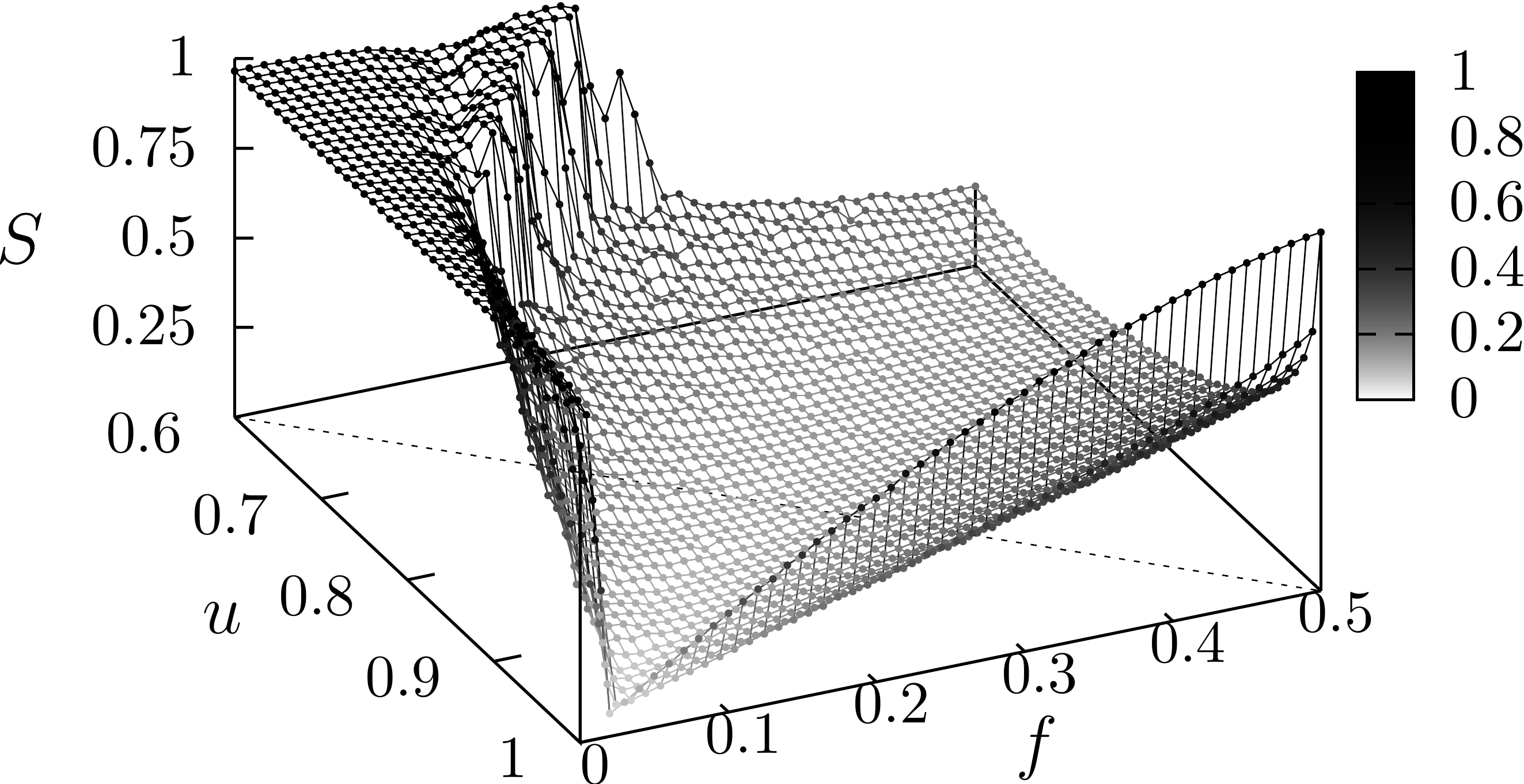} \ 
  }
\caption{Fraction of agents in the largest component of the network, $S$, in the parameters space ($u$;$f$).}
\label{fig:S}
\end{figure}

To understand the behavior of the Gini index in terms of interactions between agents, the network of agents is characterized by means of two order parameters.
The first one is the fraction of agents in the largest network component $S$,
where network component is defined as a subset of the network in which any two agents are connected to each other by paths of links.
Figure \ref{fig:S} shows the fraction of agents in the largest network component in the parameters space ($u$;$f$).
It can be appreciated a region, for small values of $f$, where $S \approx 1$, i.e., where almost all agents are interconnected on a single network component.
There is a second region of the space ($u$;$f$) where the values of $S$ are close to zero and grow smoothly when the value of $u$ increases until to reach relatively high values when $u=1$, i.e. when the concept of economic classes vanish. This means that any couple of agents that are neighboring each other can always exchange their wealths, since the inequality of eq.(\ref{dif_riquezas}) is always true.

The points $f_S(u)$, where the values of S change abruptly, define the border between these two regions.
Note that for any value of the width of the economic classes, $u$, considered in our model, the policy of redistribution of wealth may result in a fragmentation of the network of agents, and contrary to what might be expected
the value of $f_S(u)$, for wich the network is fragmented, decreases when
increases the width of the economic classes.

A second order parameter, the network modularity \cite{Newman}, is introduced to understand what occurs to the network when it goes from to be connected to fragmented.
Modularity measures the tendency of agents to group into communities or modules. Networks with a high modularity have many connections between nodes that are within the same community, but few connections between nodes that are in different communities.
The modularity is given by
\begin{equation}
Q=\frac{1}{2 N \bar{k}}\sum_{i,j}\left(a_{ij} - \frac{k_i k_j}{2 N \bar{k}}\right)
   \delta(c_i,c_j) \;,
\label{eq:Q}
\end{equation} 
where, $\bar{k}$ is the network degree; $a_{ij}$ is the $(i,j)$ component of the adjacency matrix of the network, with $a_{ij} = 1$ if $j \in \eta_i$ and $a_{ij} = 0$ if $j \notin \eta_i$; $\delta$ is the Kronecker delta; and $c_i$ is the characterizer that identify the community to which the agent $i$ belongs. In order to set the values of each characterizer $c_i $ needed to obtain the value of $Q$, the communities detection algorithm proposed by
Blondel \emph{et al.}~\cite{Blondel} is used.

\begin{figure}[htb]
\centerline{
  \includegraphics[width=0.40\linewidth,angle=90]{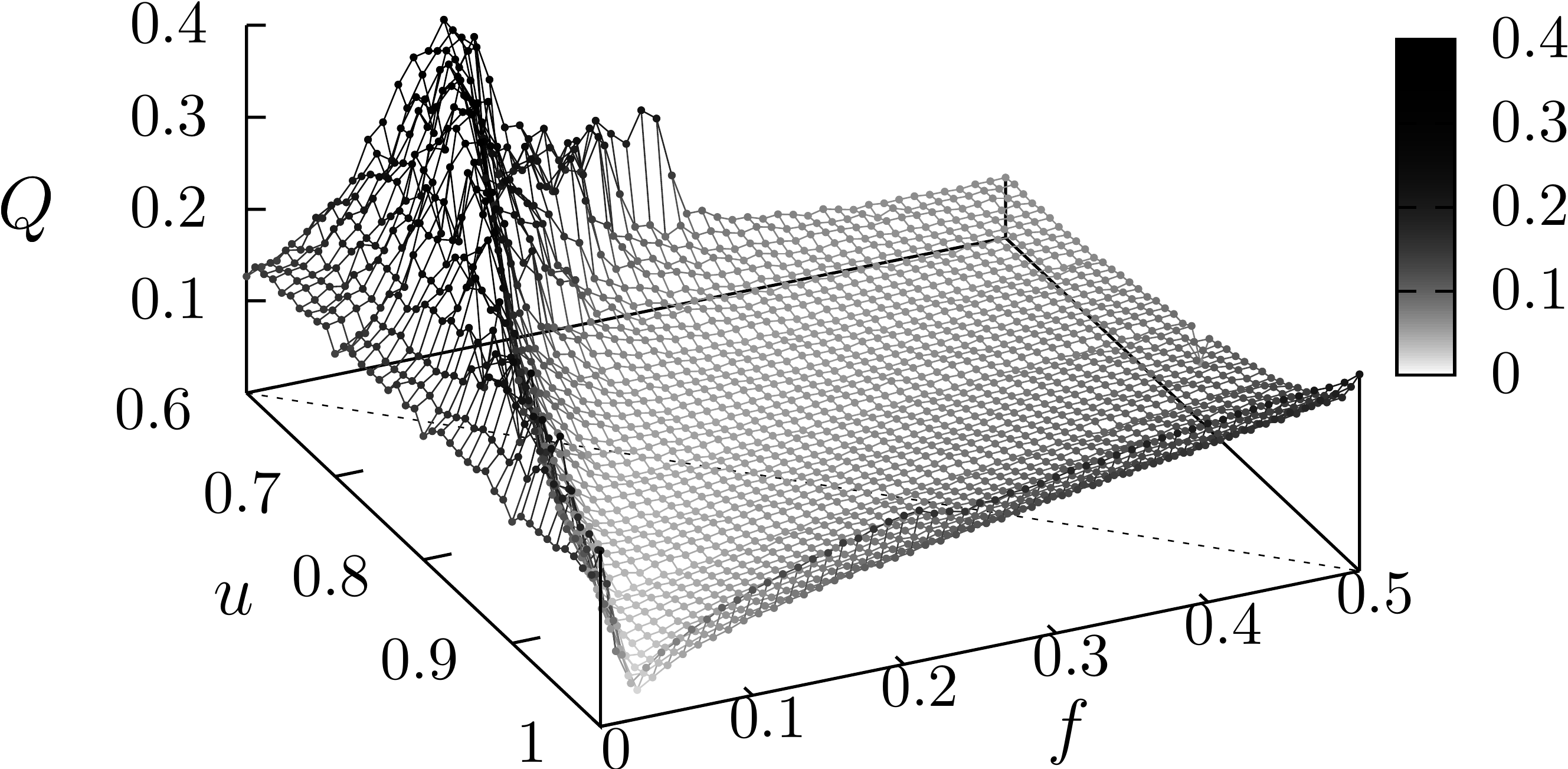}
  }
\caption{Network modularity, $Q$, in the parameters space ($u$;$f$).}
\label{fig:Q}
\end{figure}

Figure \ref{fig:Q} shows the order parameters $Q$ in the parameter space ($u$;$f$).
Similarly to figure \ref{fig:S}, this figure shows the presence of two well defined regions. The first region, which corresponds to relatively large values of modularity, has a ridgeline ($f_{Q_{max}}(u)$) and extends from the axis $u$ to the line $f_Q(u)$ where values of modularity fall suddenly.
The second region is fairly flat and runs from the border, defined by $f_Q(u)$, to the largest value of the wealth redistribution politics, $f=0.5$; and as what happens to the order parameter S, the value of Q grows smoothly when the value of $u$ increases until reach its maximum values when $u=1$.

In order to understand the relationship between the order parameters
$G$, $S$ and $Q$, the cross sections made on the diagonal of the parameters space ($u$;$f$) 
\begin{figure}[htb]
\centerline{
\includegraphics[width=0.45\linewidth,angle=90]{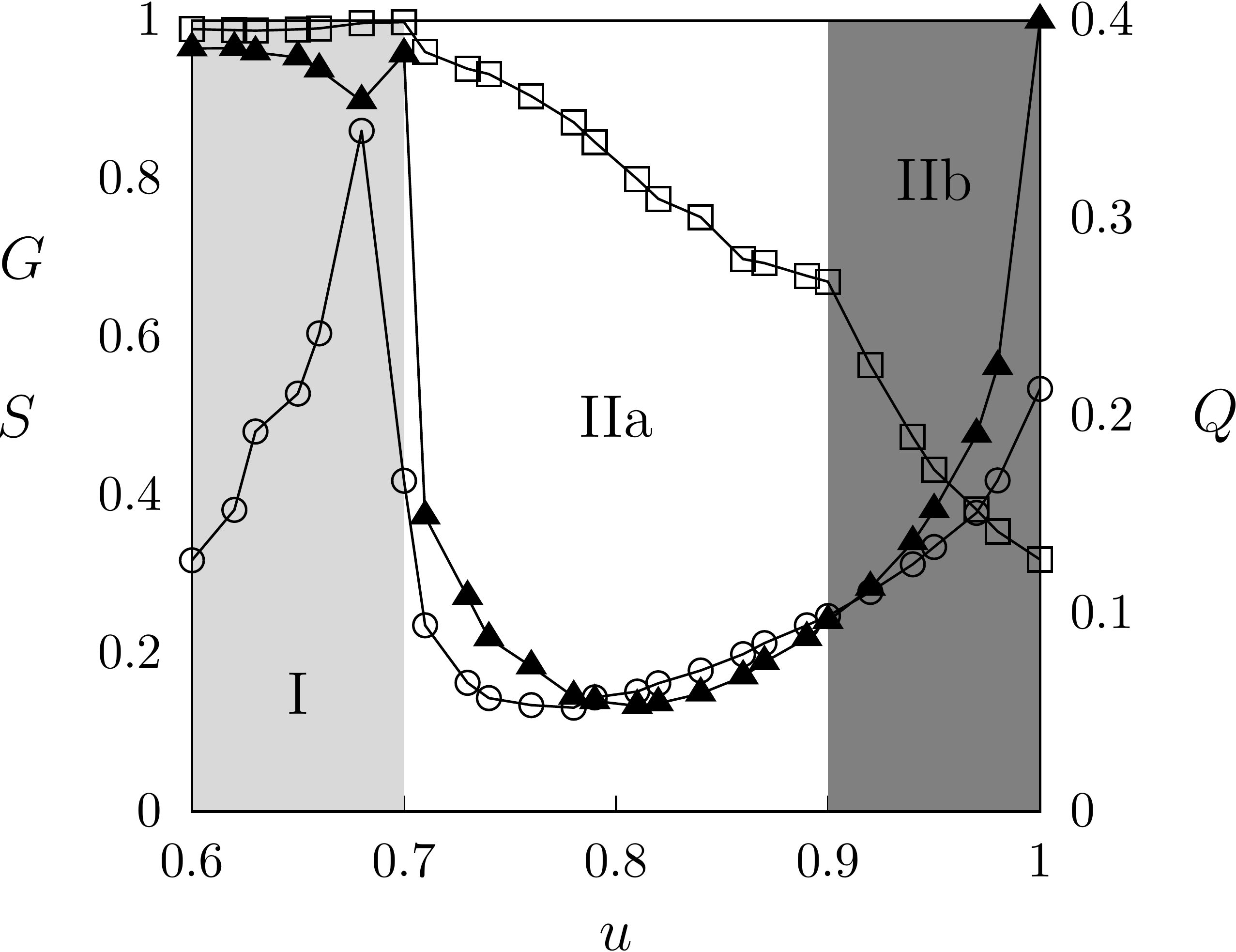}
}
\caption{Cross sections of the surfaces $G(u,f)$ (open squares), $S(u,f)$ (solid triangles) and $Q(u,f)$ (open circles); made on the diagonal of the space ($u$;$f$). 
The diagonal is given by $f(u)=5/4(u-0.6)$.
Two zones are highlighted: Absolute inequality with connected networks (light gray) and inequality similar to those observed in the world with networks that go from fragmented to connected (dark gray).
}
\label{fig:diagonal}
\end{figure} 
are shown in Figure \ref{fig:diagonal} (see dashed lines in
Figures \ref{fig:G},\ref{fig:S} and \ref{fig:Q}) with the surfaces
$G(u,f)$, $S(u,f)$ and $Q(u,f)$. 
The light gray zone (phase I) indicates the region for which all agents are connected in a single component and the wealth distribution is totally unequal.
In the other hand, parameter values $u>0.7$ (phase II) correspond to the zone where $G<1$.
At the same time, this zone can be divided into two sub-zones. In the first one,
labeled IIa, the Gini index decreases, starting from $G=1$, and networks are fragmented. 
Note that the value of the parameter $u$ where the Gini index value begins to decrease is the same value for which the fraction of agents in the largest component and the modularity of the network fall abruptly.
In other words, when the width of the economic classes $u$ increases (in this figure the redistribution politics $f$ also increases)
there is a phase transition
from a totally unequal to a more equitable distribution of wealth, achieved at the expense of the fragmentation of the economic network.
The second sub-zone (IIb) corresponds to the region where the Gini index reaches values that fit to those observed in most countries $(G \in [0.2;0.70])$ and the world as a whole, that has been estimated  $G \in [0.60;0.63]$~\cite{Milanovic}.
In this region we can see how when the width of the economic classes increases, the Gini index decreases, while the modularity increases as well as it  does the fraction of agents in the largest component which reaches up to $S=1$ when $u=1$.

\begin{figure}[htb]
\centerline{
\includegraphics[width=0.50\linewidth,angle=90]{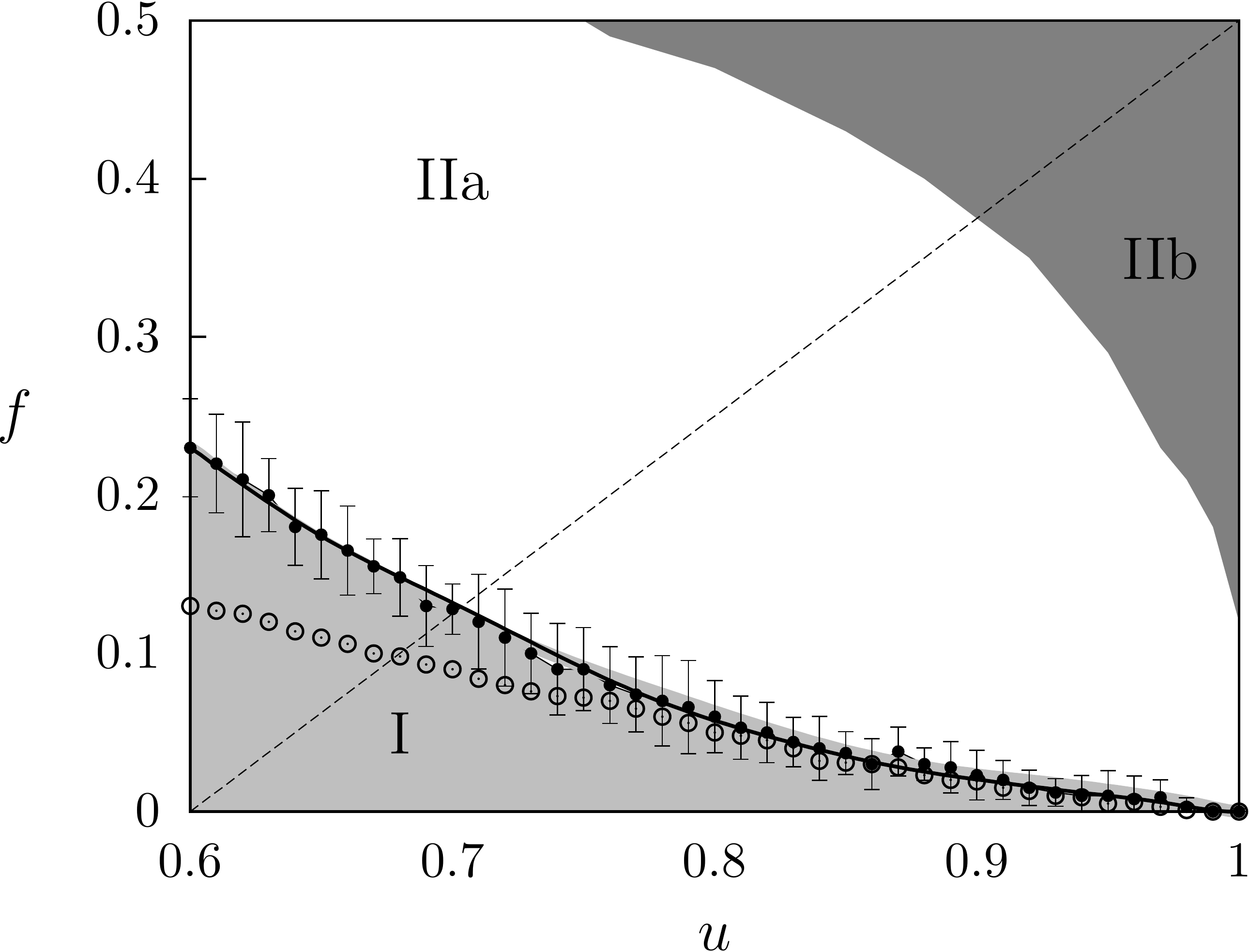}
}
\caption{Phase diagram in the parameters space ($u$;$f$).
The critical boundary between phases I and II can be established either by the critical values with any of the three order parameters: $f_G$ (solid line) $f_S$ (border of the light gray area) or $f_Q$ (solid circles).
Open circles represent the ridgeline values of the modularity,  $f_{Q_{max}}$.
The region in which the Gini index is consistent with the values observed in the world is shown in dark gray (IIb).
Dashed line correspond to the cross section used to Figure \ref{fig:diagonal}.
}
\label{fig:fase}
\end{figure}

To  characterize the statistical properties of the system, in Figure \ref{fig:fase} it is shown the phase diagram of the system on the parameters space ($u$;$f$).
There are two zones where modularity is significantly high,
revealing the presence of communities of agents in the network large enough to be detected by the algorithm proposed by Blondel et al. \cite{Blondel}.

\begin{figure}[htb]
\centerline{
\includegraphics[width=0.35\linewidth,]{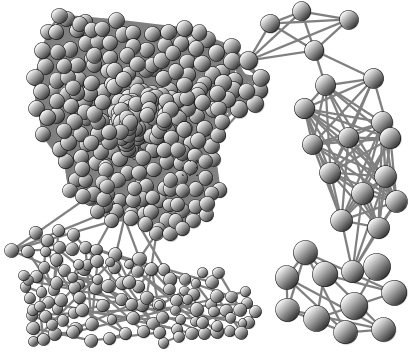}
\hspace{10mm}
\includegraphics[width=0.35\linewidth]{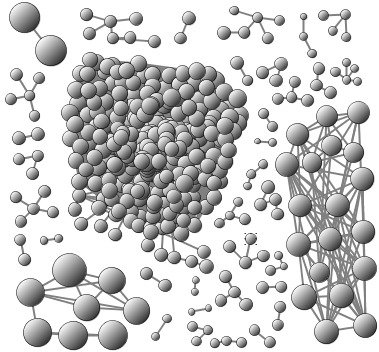}
}
\caption{Snapshot of the structure of networks of active agents for two different values of parameters in phase I.
Left: $u=0.68$ and $f=0.1$, correspond to the point where $Q$ reaches its maximum in Figure \ref{fig:diagonal}.
Right: Parameters $u=0.71$ and $f=0.125$ are set on the critical point between phases I and II.
Nodes size represents the agents wealth in logarithmic scale.
Snapshot are done with the help of Network Workbench~\cite{NWB}.
}
\label{fig:faseI}       
\end{figure}

The first zone matches with the phase I, where, even though the order parameters $G$ and $S$ do not change significantly, the value of $Q$ does, reaching a maximum and then decaying before reaching the critical boundary, $f_Q$.
These changes in the values of $Q$ indicate an agents rearrangement with the subsequent formation of several communities on the network when the parameters $f$ and $u$ change.
Figure \ref{fig:faseI} shows two snapshot of the structure that are formed in the network of agents for two different points in the parameter space ($u$;$f$).
In both snapshot are shown active links, i.e. those for which eq.(\ref{dif_riquezas}) is true, and agents connected through these links (approximately 500 active agents). 
The wealth of each agent is represented by its size in logarithmic scale.

It can be appreciated in the left snapshot, which corresponds to the diagonal point in the plane ($u$;$f$) where $Q$ is maximum,
a connected network with five communities of active agents with similar wealth.
In contrast, in the right snapshot, which corresponds to the point of transition between phases I and II, the network of active agents is no longer connected, but the modularity is maintained relatively high by the presence of unjoined communities. 
At this point, where the communities disappear or become smaller and disconnected, the random structure of the network of inactive agents is imposed and therefor the modularity falls, the network gets fragmented and exchange of wealth declines.

In the phase II, characterized by a decrease in the Gini index when the value of the parameters ($f$, $u$  or both) increases, network fragmentation $S$ and modularity $Q$ achieve their minimum values, as shown in Figure \ref{fig:diagonal}.
Beyond this point, by increasing the values of $f$ or $u$, both order parameters $S$ and $Q$ grow indicating that agents have been rearranged and creating structures within the network.
This process continues until connected networks with relatively high levels of modularity, $Q > 0.1$, emerge around the point ($u$;$f)~=~(1.0;0.5)$. 

\begin{figure}[htb]
\centerline{
\includegraphics[width=0.60\linewidth,]{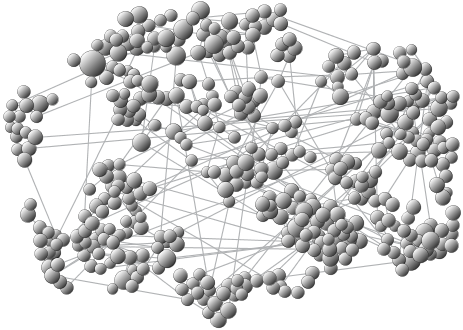}
}
\caption{Snapshot of the network structure network for $u=1.0$ and $f=0.5$ 
Network size $N=500$ and degree $\bar{k}=4$.
Nodes size represents the agents wealth in logarithmic scale.
}
\label{fig:redIIb}       
\end{figure}

Note that at this point all links are active since eq. (\ref{dif_riquezas}) is true for all pairs of neighbors.
In other words, networks of active agents are composed by $N = 10^4$ agents linked together through $N \times \bar{k}=160000$ links, all active. 
performing a visual inspection of a network of this size in order to observe communities is not possible, therefore, to see the structures that arise in these conditions, Figure \ref{fig:redIIb} shows a network with $\bar{k}=4$ and $N=500$ agents, that is a similar size of the two networks shown in Figure \ref{fig:faseI}.
The presence of communities can be clearly appreciated in the figure although they are not as well defined as the communities shown in Figure 6.
It may also be noted that in this case the diversity of wealth of agents (size of nodes) is larger within each community. However, if considering the system as a whole, the wealth diversity is smaller, which is consistent with the values obtained for the Gini index at this point.

\section{Discussion}
\label{sec:Discussion}

Using the concepts of distribution of wealth among agents with risk aversion \cite{Iglesias}, economic stratification \cite{Laguna} and spatially localized interactions \cite{JL}, a wealth exchanging multi-agent model was implemented where agents can also change their environment, i.e. a co-evolutionary dynamic system where agents can change their neighborhood and these topological changes have effects on the dynamics of the agents.
The model can fit into the general framework for systems with coevolution between topology and dynamics \cite{HCTG2011} as a DR process, i.e. a process with a rewiring dynamic where disconnect actions are governed by a dissimilarity mechanism (D) and reconnection actions are governed by a random mechanism (R).

As might be expected, increased redistribution policy of the government, represented by the parameter $f$, achieves a more equitable distribution of wealth, i.e., Gini index, $G$, decreases. As is shown in Figure \ref{fig:fase}, this behavior allows to define two phases in the parameter space ($u$;$f$): Phase I, where $ G \approx 1$, and phase II, where $ G < 1$. 
Is appreciated that the critical boundary between phases I and II, $f_G(u)$, depends on the parameter that controls the width of the economic classes, $u$. As the economic classes vanish a less aggressive redistribution policy is required for the transition from phase I to phase II.

Observing the behavior of the order parameters $S$ and $Q$, both related to the structure of the network and not with the wealth of agents such as $G$, highlights the fact that just at the critical boundary their values change significantly which leads to conclude that the phase transition of the system is related to the network fragmentation and the vanishing of community structure in it, characterized through $S$ and $Q$ respectively.

Finally, the results show two zones in the phase diagram of the system where communities of agents emerge spontaneously. The first zone coincides with the phase I of system where there is maximum inequity in the distribution of wealth. The communities in this zone, composed of elements with similar wealth, are chained one after another sorted by the richness of agents.

The second zone where communities emerge is in Phase II, specifically within the region where the Gini index fits the values currently observed in the world.
Here, unlike what is seen in the previous zone, the variety of wealth of the agents that compose each community is large and communities seem to have no order within the network.

In this region of the parameters space the coevolutionary model of wealth exchange proposed  gives as results connected networks in which spontaneously emerge communities of agents with diverse wealths and Gini index values similar to those seen currently in the world.

\section{Acknowledgements}
This work is supported partly by Grant No. C-1804-12-05-B CDCHTA, Universidad de Los Andes, Venezuela.


\begin{thebibliography}{}
\bibitem{epidemic1} 
S. Yu-Rong, J. Guo-Pin and G. Yong-Wang, Chinese Physics B \textbf{22}, (2013) 040205. 

\bibitem{epidemic2} 
S. Wieland, T. Aquino and A. Nunes, Europhysics Letters \textbf{97}, (2012) 18003.

\bibitem{technicalNet1} 
M. Lim, D. Braha, S. Wijesinghe, S. Tucker and Y. Baryam, Europhysics Letters \textbf{79}, (2007) 58005-6.

\bibitem{technicalNet2} 
A. Scir\`e, I. Tuval and V. M. Egu\'iluz, Europhysics Letters \textbf{71}, (2005) 318-424.

\bibitem{neural1} 
D. Moriarty and R. Miikkulainen, Evolutionary Computation \textbf{5}, (1997) 373-399.

\bibitem{neural2} 
J. Ito and K. Kaneko, Neural Networks, \textbf{13}, (2000) 275.

\bibitem{social1} 
P. Holme and M. E. J. Newman, Physical Review E, \textbf{74}, (2006) 056108.

\bibitem{social2} 
F. Vazquez, J. C. Gonz\'alez-Avella, V. M. Egu\'iluz and M. San Miguel, Physical Review E, \textbf{76}, (2007) 46120.

\bibitem{social3} 
D. Kimura and Y. Hayakawa, Physical Review E, \textbf{78}, (2008) 016103.

\bibitem{game} 
B. Skyrms, and R. Pemantle, PNAS, \textbf{97}, (2000) 9340-9346.

\bibitem{eco1} 
U. Dieckmann and M. Doebeli, Nature \textbf{400}, (1999) 354-357.

\bibitem{eco2} 
J. L. Garcia-Domingo and J. Salda\~na, Journal of Theoretical Biology \textbf{247}, (2007) 819-826.

\bibitem{chemical1} 
S. Jain and S. Krishna, PNAS \textbf{98}, (2001) 543-547. 

\bibitem{chemical2} 
A. M. Seufert and F. Schweitzer, International Journal of Modern Physics \textbf{18}, (2007) 1-18.

\bibitem{HCTG2011} 
J.L. Herrera, M.G. Cosenza, K. Tucci and J.C. González-Avella, Europhysics Letters \textbf{95}, (2011) 58006.

\bibitem{PI03} 
S. Pianegonda, J.R. Iglesias, G. Abramson and J.L. Vega, Physica A \textbf{322}, (2003) 667-675.

\bibitem{Iglesias} 
J.R. Iglesias, S. Goncalves, G. Abramson and J.L. Vega, Physica A \textbf{342}, (2004) 186–192. 

\bibitem{Laguna} 
M.F. Laguna, S.R. Gusman and J.R. Iglesias, Physica A \textbf{356}, (2005) 107–113.

\bibitem{JL} 
J. L. Herrera, M. G. Cosenza and K. Tucci, Revista Cient\'ifica UNET \textbf{21}, (2009) 8. 

\bibitem{HCT2011} 
J. L. Herrera, M. G. Cosenza and K. Tucci, Physica A \textbf{390}, (2011) 1453–1457.

\bibitem{IglesiasGoncalvesPianegonda} 
J.R. Iglesias, S. Goncalves, S. Pianegonda, J.L. Vega and G. Abramson, Physica A \textbf{327}, (2003) 10002-10017.

\bibitem{Garlaschelli} 
D. Garlaschelli and M. Loffredo, Journal of Physics A: Mathematical and Theoretical \textbf{41}, (2008) 224018. 

\bibitem{ER59} 
P. Erd\"os and A. R\'enyi, Publicationes Mathematicae \textbf{6}, (1959) 290-297.

\bibitem{Newman} 
M. E. J. Newman, PNAS {\bf 103}, (2006) 8577-8696.

\bibitem{Blondel} 
V. D. Blondel, J. L. Guillaume, R Lambiotte, and E. Lefebvre, Journal of Statistical Mechanics \textbf{10}, (2008) 10008.

\bibitem{Milanovic} 
B, Milanovic, Policy Research Working Paper, World Bank, {\bf 5044} (2009). 

\bibitem{NWB} 
NWB Team,
Indiana University, Northeastern University, and University of Michigan, http://nwb.slis.indiana.edu. (2006).

\end{thebibliography}
\end{document}